\documentclass[fleqn,10pt]{wlscirep}
\usepackage[utf8]{inputenc}
\usepackage[T1]{fontenc}
\usepackage{float}
\usepackage{CJKutf8} 
\usepackage{ulem}
\usepackage{lineno}  
\usepackage{longtable}
\usepackage{newtxtext,newtxmath}
\usepackage{jabbrv}

\addto\captionsenglish{}
\title{Interpretable structural–semantic decoding reveals language-like organisation of regulatory information in DNA}

\author[]{Li Yang}
\author[1,*]{Dongbo Wang}
\affil[]{School of Information Management, Nanjing Agricultural University, Nanjing, 210095, China}

\affil[*]{corresponding author @ db.wang@njau.edu.cn (D. Wang)}

\begin{abstract}
Decoding how linear DNA encodes regulatory information remains a central challenge. Existing decoding approaches lack interpretability and struggle to reveal the underlying coding principles. Here, we present the interpretability-first, structural artificial intelligence (AI) framework for DNA (ISAF4DNA), which uses state-aware symbolic encoding and couples structural unit discovery with semantic validation to form a closed-loop structural–semantic decoder. When applied to N6-methyladenine (6mA) datasets from 63 species, ISAF4DNA reveals a language-like organization of regulatory information: (i) a conserved motif-derivation pathway AT$\rightarrow$GAT/ATC$\rightarrow$GATC; (ii) two forms of redundant syntax—anchor-type structures with conserved core and selective flanks and fuzzy-type clusters composed of distributed units with positional tolerance; and (iii) differential deployment trends between prokaryotes and multicellular eukaryotes. Together, these observations motivate the development of a testable framework, EpigenoLinguistics, that treats motifs as lexical units, redundancy as syntax, and deployment as pragmatics. This framework advances the “DNA as language” concept from a metaphor to a falsifiable framework with supporting evidence, thereby bridging biology and computational linguistics. ISAF4DNA advances the application of AI techniques in biology from black-box predictions to mechanism-level signals, augments database annotations, and guides regulatory-element design, with principles extensible to other modifications.

\end{abstract}

\begin{document}

\flushbottom
\maketitle
%
%

\begin{figure}[H]
	\centering
	\includegraphics[width=0.70\textwidth]{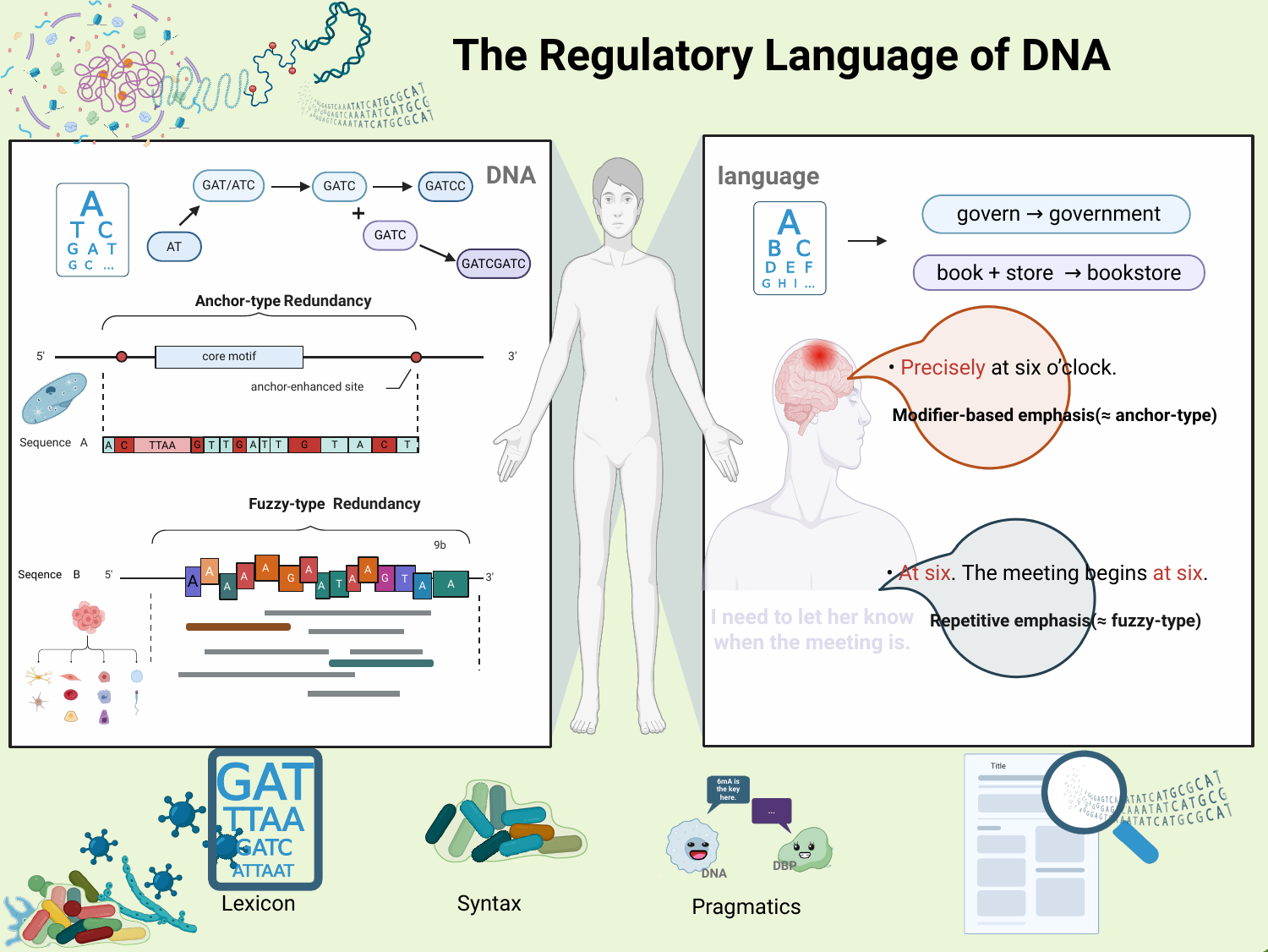}
	\caption*{\textbf{Graphical Abstract | ISAF4DNA reveals the principles of DNA regulatory grammar and advances a testable EpigenoLinguistics framework.} 
	}
\end{figure}
\thispagestyle{empty}
\section*{Introduction}

In the post-genomic era, the acquisition of DNA sequences has become routine. The true challenge lies in deciphering their underlying coding logic—a central question to understanding life’s information system and enabling its programmable control. In recent years, AI techniques have become increasingly integrated into the life sciences discovery pipeline, moving beyond analytical enhancement to hypothesis generation, experimental design optimization, the interpretation of complex, large-scale datasets, and the extraction of insights unattainable by conventional approaches\cite{1}. Such integration represents more than a methodological upgrade—it opens an unprecedented route to decode the higher-order organizational logic embedded within DNA’s coding logic.

Although AI technologies continue to expand methodological possibilities and numerous natural language processing (NLP) algorithms have been applied directly to DNA sequence analysis\cite{2,3,4}, the fundamental question—whether DNA regulatory systems genuinely embody information-organizational principles analogous to those of natural language—remains systematically unexplored. For decades, discourse on the linguistic properties of DNA has remained largely metaphorical\cite{5,6,7}and often contentious\cite{8}. As a result, a rigorous ontological framework is critically absent.

A range of DNA decoding strategies have emerged in recent years, particularly supervised deep predictive models\cite{9}and unsupervised DNA large language models\cite{10,11,12}, in addition to long-established motif discovery tools\cite{13,14}. Despite advances in predictive accuracy, distributed representation learning, and motif discovery capabilities, these methods often exhibit limitations—namely, structural opacity, limited interpretability, and a lack of context integration\cite{1,15}—that hinder the recovery of the regulatory organization of DNA. The root cause lies in a prevailing simplification: most current approaches conceptualize DNA as a static, linear code. This view overlooks the role of DNA as a dynamic information-processing system\cite{16,17}shaped by structural state, local context, and enzymatic recognition, thereby limiting our ability to reconstruct the organizational logic of regulation. Systems biology emphasizes that fully understanding biological regulation involves looking beyond static snapshots of genes and proteins to reveal the structural and dynamic principles that shape system-wide behaviour\cite{18}. From an information-theoretic perspective, life itself can be viewed as a multilayered information-processing system, with genetic material serving as the active medium for acquiring, transforming, storing, and orchestrating information\cite{19}. Within this framework, decoding the organizational principles of DNA is not optional—it is foundational to unifying the sequence structure with regulatory logic and functional meaning. Nonetheless, prevailing decoding strategies remain limited in interpretability and overlook context, which leaves the core architecture of DNA’s regulatory grammar unresolved.

To address this gap, we propose ISAF4DNA, a mechanism-transparent, dual-layer interpretable AI framework that systematically connects structural recognition, semantic evaluation, and mechanistic modelling. At its core, ISAF4DNA employs a state-aware, high-entropy symbolic encoding system that emulates the structural dynamics of enzyme–DNA recognition, thereby substantially increasing the informational expressiveness of individual nucleotide sites. By integrating structural rule extraction with semantic consistency verification, ISAF4DNA establishes a closed-loop decoding pipeline. This pipeline progresses from structural recognition to semantic evaluation and mechanistic modelling, enabling robust, cross-species and cross-model interpretation of regulatory syntax. In this way, ISAF4DNA effectively overcomes the structural opacity and context neglect prevalent in existing approaches.

When applied to 6mA data from 63 species and without linguistic priors, ISAF4DNA delineates a three-layer information architecture—lexicon, syntax and pragmatics—and operationalizes the “DNA as language” concept into an empirically testable framework, EpigenoLinguistics. The approach reconstructs motif-level derivation pathways and redundancy-aware syntactic patterns, refines and extends community databases (e.g., REBASE), and provides an experiment-ready 6mA coding catalogue, together with parameterized sequence-design templates. Collectively, these advances—along with open-access 6mA resources—shift AI4Science from black-box prediction to interpretable, mechanism-level insight and lay a practical foundation for reading, rewriting, and engineering regulatory codes.

\section*{Results}


\subsection*{Overview of ISAF4DNA}

\begin{figure}[H]
	\centering
	\includegraphics[width=1.0\textwidth]{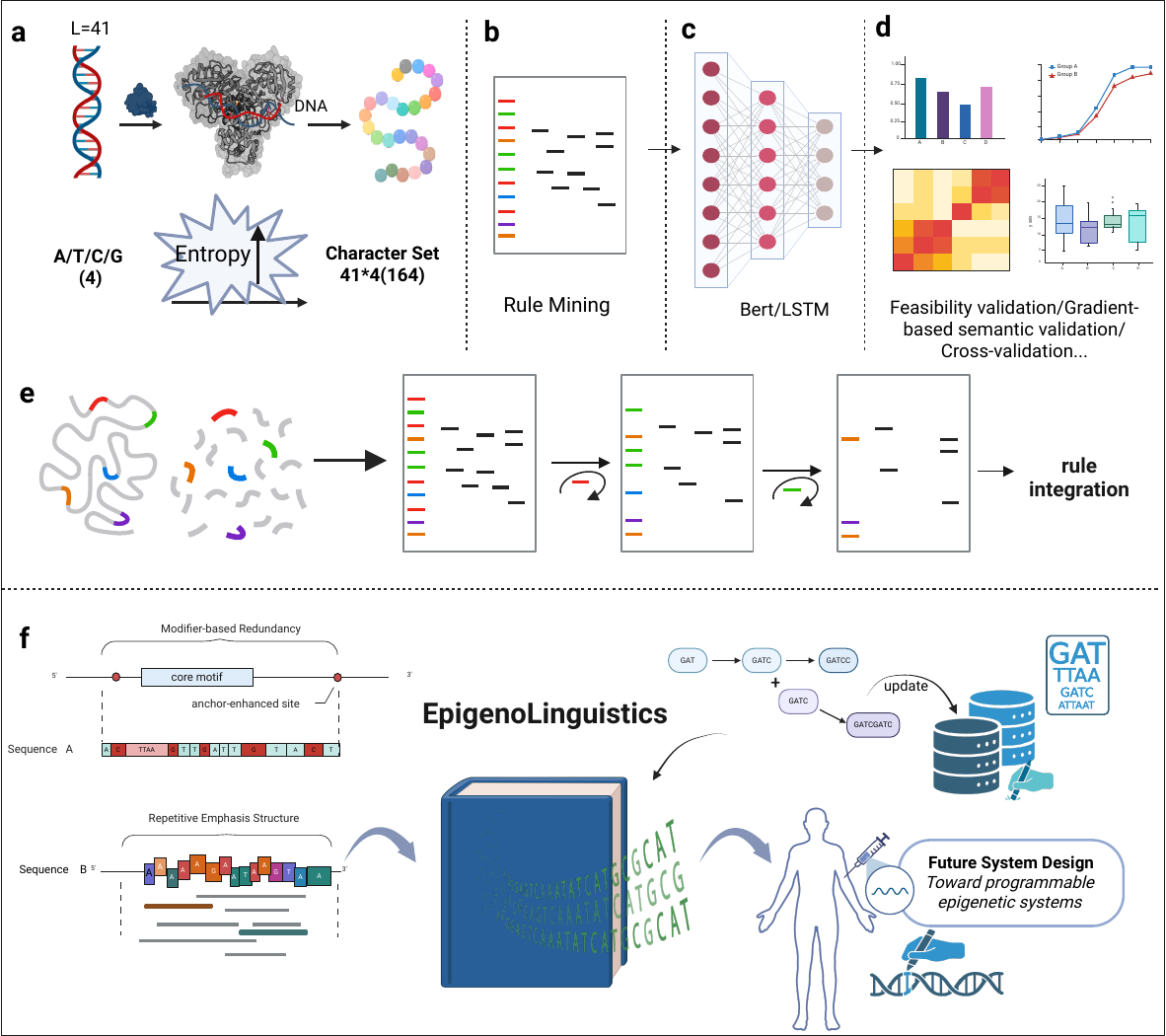}
	\caption{\textbf{The ISAF4DNA framework for decoding DNA regulatory grammar.}(a) State-aware symbolic encoding abstracts the biophysical features of enzyme–DNA recognition into discrete states, which allows high-capacity sequence representation. (b) The discovery of reusable structural units from high-entropy symbolic sequences builds a foundational library for regulatory grammar. (c) AI-driven evaluation of structure–function consistency assesses the regulatory rules inferred from structural units. (d) A multidimensional validation of the structural units assesses the functional relevance, cross-species generalization, and consistency with experimental data. (e) Within-species completion of structural-unit–based rules maps function-specific structural patterns in DNA. (f) The application of the framework reveals two distinct categories of redundant structures, novel motifs, and site-shifting mechanisms. These findings support a language-like organization of DNA regulation and inform the expansion of functional databases and the design of programmable epigenetic systems. The figure was created with BioRender.com.} 
	\label{fig1} 
\end{figure}

ISAF4DNA implements a state-aware high-entropy symbolic encoding framework whereby the biophysical features that underlie enzyme–DNA recognition (e.g., hydrogen bonding patterns and conformational angles) are abstracted as additional state dimensions. This abstraction expands the nucleotide alphabet from 4 to 164 symbols (Fig. \ref{fig1}a), which enables a maximum entropy of 7.36 bits per character (Methods)—a 3.68-fold increase in information capacity compared with that of canonical four-base encoding. In this way, ISAF4DNA transcends the constraints of conventional nucleotide representation and establishes a pathway towards mechanistic transparency. The framework couples three complementary discovery modules—structural unit recognition (Fig. \ref{fig1}b), functional consistency assessment (Fig. \ref{fig1}c), and hierarchical annotation with completion (Fig. \ref{fig1}e)—into a closed-loop structure–function decoding system. A multidimensional validation architecture (Fig. \ref{fig1}d) ensures both the stability and cross-species adaptability of the resulting rules.

When applied to 6mA datasets that span 63 species from bacteria to eukaryotes, ISAF4DNA reveals language-like regularities in DNA regulation. We identify two categories of redundant nested structures that have not, to our knowledge, been systematically characterized across species, and chart derivational pathways within motif families. These findings empirically support viewing DNA as a dynamic and context-dependent regulatory system. On this basis, we introduce EpigenoLinguistics, a cross-species framework that integrates structural redundancy, functional grammar and linguistic analogy. Our work lays a basis for regulatory-grammar decoding and—in principle—for extension to other modification types; it also supports the augmentation and calibration of databases, and informs the rational design of programmable regulatory systems (Fig. \ref{fig1}f).

\subsection*{Encoding Layer: Design and Validation of the State-Aware Symbolic System}
\begin{figure}[H]
	\centering
	\includegraphics[width=1.0\textwidth]{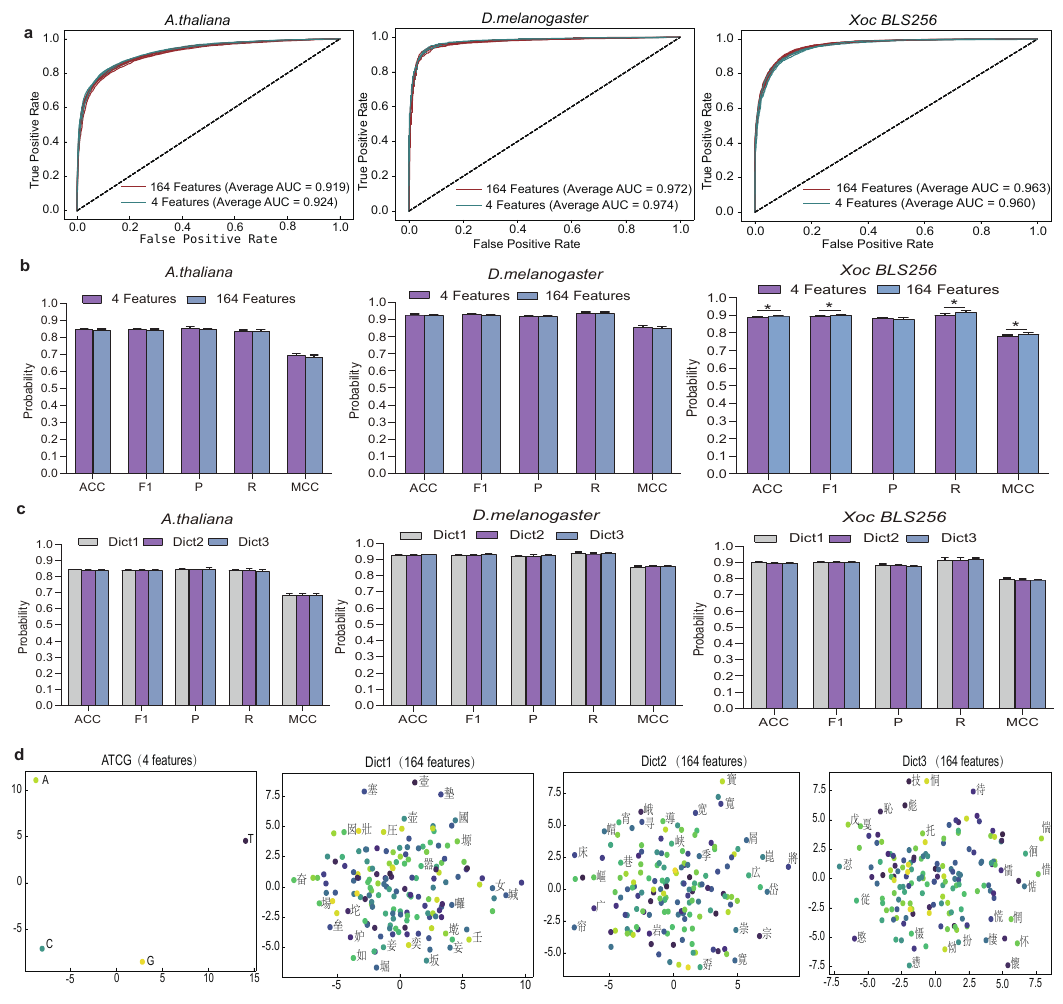}
	\caption{\textbf{The encoding layer: A comparison of state-aware symbolic encoding with conventional nucleotide encoding for DNA 6mA site prediction.}(a) ROC curves from a 5-fold cross-validation comparing 4-feature (A/T/C/G) and 164-feature (state-aware symbolic encoding) inputs using the same pretrained BERT model, with their average AUC. (b) Performance comparison of the 4-feature and 164-feature models in terms of accuracy (ACC), precision (P), recall (R), F1, and Matthews correlation coefficient (MCC). (c) Evaluation of 164-feature encodings with different dictionaries (Dict1, Dict2, and Dict3) are evaluated in terms of ACC, F1, precision (P), recall (R), and MCC. (d) Two-dimensional PCA projection of embeddings extracted from the pretrained BERT model. Statistical significance was assessed using a two-sided Student’s t test; only significant differences are indicated ($\ast P < 0.05$; $\ast\ast P < 0.01$).} 
	\label{fig2} 
\end{figure}

To establish the validity and generality of the state-aware symbolic encoding system, we benchmarked its performance on the basis of three representative species—\textit{A.thaliana} (plant), \textit{D.melanogaster} (animal), and \textit{Xanthomonas oryzae} pv. Oryzicola BLS256 (\textit{Xoc BLS256}) (microorganism).

The state-aware symbolic encoding representation demonstrated strong performance within pre-trained language models (BERT). When BERT was trained   based on symbolic sequences, the model achieved robust classification of 6mA sites across all species (Fig. \ref{fig2}a), with AUC values exceeding 0.90 and peaking at 0.97 in \textit{D.melanogaster}.

The performance was comparable with—or marginally superior to—the canonical four-base representation. The models trained with state-aware symbolic features exhibited predictive performance metrics  (ACC, P, R, F1, and MCC) comparable to those trained with conventional A/T/C/G features (Fig. \ref{fig2}b). In \textit{A.thaliana} and \textit{D.melanogaster}, no statistically significant differences were observed, whereas in Xoc BLS256, symbolic encoding conferred modest yet consistent improvements (Fig. \ref{fig2}b). These results suggest that the state-aware symbolic encoding system captures species-specific informational nuances without compromising predictive power.

The encoding scheme itself was robust and semantically neutral. We validated this finding by constructing three disjointed symbolic mapping dictionaries (Fig. \ref{fig2}d). Across all of the models trained based on these distinct dictionaries, the predictive outcomes remained statistically indistinguishable (Fig. \ref{fig2}c). This finding demonstrates that the effectiveness of the state-aware symbolic encoding system arises from its representational logic rather than from the particular symbol assignments, which underscores its robustness, generality, and theoretical soundness.

\subsection*{Spontaneous Emergence of Redundant Structures under Structural–Functional Dual-Layer Decoding}

At the structural layer, frequent-pattern analysis revealed structural units markedly enriched in positive samples across three species relative to the negative controls (Supplementary Tables 3–5). These structural units ($P(\text{pos}) \geq 0.90$) aggregated into overlapping and nested clusters within sequences (Supplementary Fig. 2) and formed redundant patterns consistently observed in \textit{A. thaliana},  \textit{D. melanogaster}, and  \textit{Xoc BLS256} (Supplementary Data 1–3). This redundancy stabilized methyltransferase recognition, thereby conferring robustness to information transmission even under mutational or conformational perturbations. Importantly, these architectures were not imposed by the model but emerged spontaneously during unsupervised decoding, which reveals a fundamental principle of molecular information organization. Their highly nested and recurrent distributional features parallel syntactic patterns in natural language and offer mechanistic clues to the organisation of DNA information(see Supplementary Fig. 2 and Fig. \ref{fig5}).

\begin{figure}[H]
	\centering
	\includegraphics[width=1\textwidth]{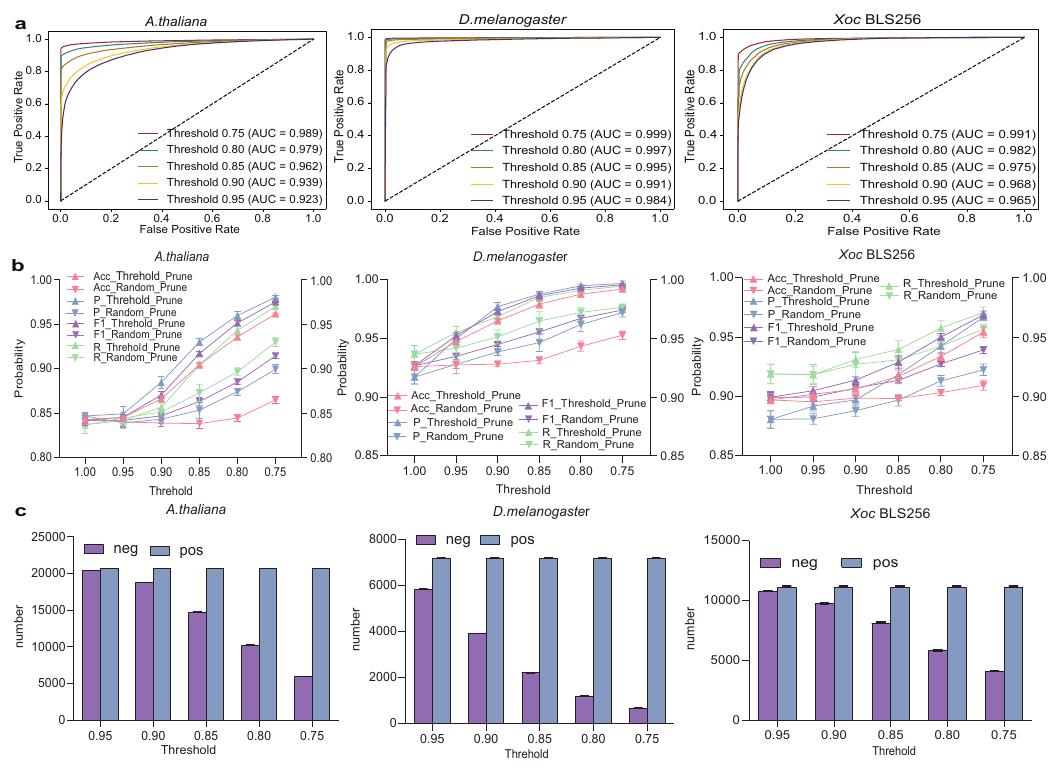}
	\caption{\textbf{Structure–function causality test: Evidence for structural units' causal role in 6mA regulation. },(a) Structural units decoded by ISAF4DNA were used to eliminate potential false negatives at different thresholds (P(pos) $\in$ {0.95, 0.90, 0.85, 0.80, 0.75}), and their impact was assessed in a BERT model with 164 feature inputs. The ROC curves and mean AUC values were obtained. Here, P(pos) represents the relative enrichment of a structural unit in positive samples, which is defined as the proportion of its frequency in positives to its overall frequency in the dataset. (b) Under the same thresholds, we compared BERT performance between the threshold-based elimination of potential false negatives and random deletion (control) using ACC, F1, P, and R as evaluation metrics. (c) The changes in the number of positive and negative samples under different threshold conditions are reported.} 
	\label{fig3} 
\end{figure}

At the functional validation level, we designed a gradient-pruning experiment by setting five thresholds (0.95, 0.90, 0.85, 0.80, and 0.75) on the positive sample proportion (P(pos)) of structural units, and we selectively pruned potential false negatives. The results confirmed causal contributions (Fig. \ref{fig3}a–b): after pruning, the BERT model showed systematic performance gains, with the AUC approaching the theoretical limit at a threshold of 0.75 (\textit{A. thaliana}: 0.99; \textit{D. melanogaster}: 1.00; and\textit{Xoc BLS256}: 0.99; Fig. \ref{fig3}a). ACC, F1, P, and R were markedly improved, and the performance was significantly better than that with the random-pruning controls (Fig. \ref{fig3}b). Moreover, the performance gains were independent of changes in the sample proportion (Fig. \ref{fig3}c), demonstrating that they arose from the intrinsic discriminative power of structural units. These findings indicate that context-aware symbolic encoding enhances structural separability, enabling systematic recognition of redundant patterns and confers decisive discriminative power in functional validation.

Cross-species validation (Supplementary Fig. 3a) identified the differential features of regulatory “syntax” among species. Although the frequency distributions of \textit{A. thaliana} and  \textit{D. melanogaster} were highly similar (Supplementary Fig. 3b–c), their underlying contribution patterns differed. For example, in the similar motif pattern (AAAGA), fragments in A. thaliana were generally shorter, suggesting functional relatedness while reflecting species-specific structural differences. By leveraging its dual-layer decoding system, ISAF4DNA captured these subtle differences, thereby providing an interpretable perspective on DNA coding mechanisms and enabling rigorous cross-species comparisons.

\subsection*{Applicability and Generalizability of ISAF4DNA: Cross-Species, Cross-Model, and External Validation}

After establishing the effectiveness of the structural–semantic decoding framework, we evaluated its applicability and generalizability along three axes: (i) whether structural unit-driven data optimization improves classification performance; (ii) whether such gains reproduce across species, model architectures, and input representations (63 species; BERT and LSTM; reverting to a 4-feature A/T/C/G encoding); and (iii) whether the computational findings align with databases and experiments and yield incremental annotations. Overall, structural unit–guided optimization improved ACC/AUC/F1/MCC and remained advantageous for the independent 6mA-DNA-IP-seq datasets. The gains held for both the BERT and LSTM models and under the 4-feature A/T/C/G encoding, which indicated noise suppression (false-negative cleaning) rather than architecture- or language-prior specific effects. The findings using ISAF4DNA agreed with those obtained using REBASE, structural‐biology evidence and added/corrected motif and site-shift annotations. Collectively, these results establish these structural units as portable and testable, with conserved signals, enabling database refinement and guiding downstream experimental design.

\subsubsection*{Structural unit-driven data curation improves performance: Baseline comparisons and external validation}

\begin{figure}[H]
	\centering
	\includegraphics[width=\textwidth]{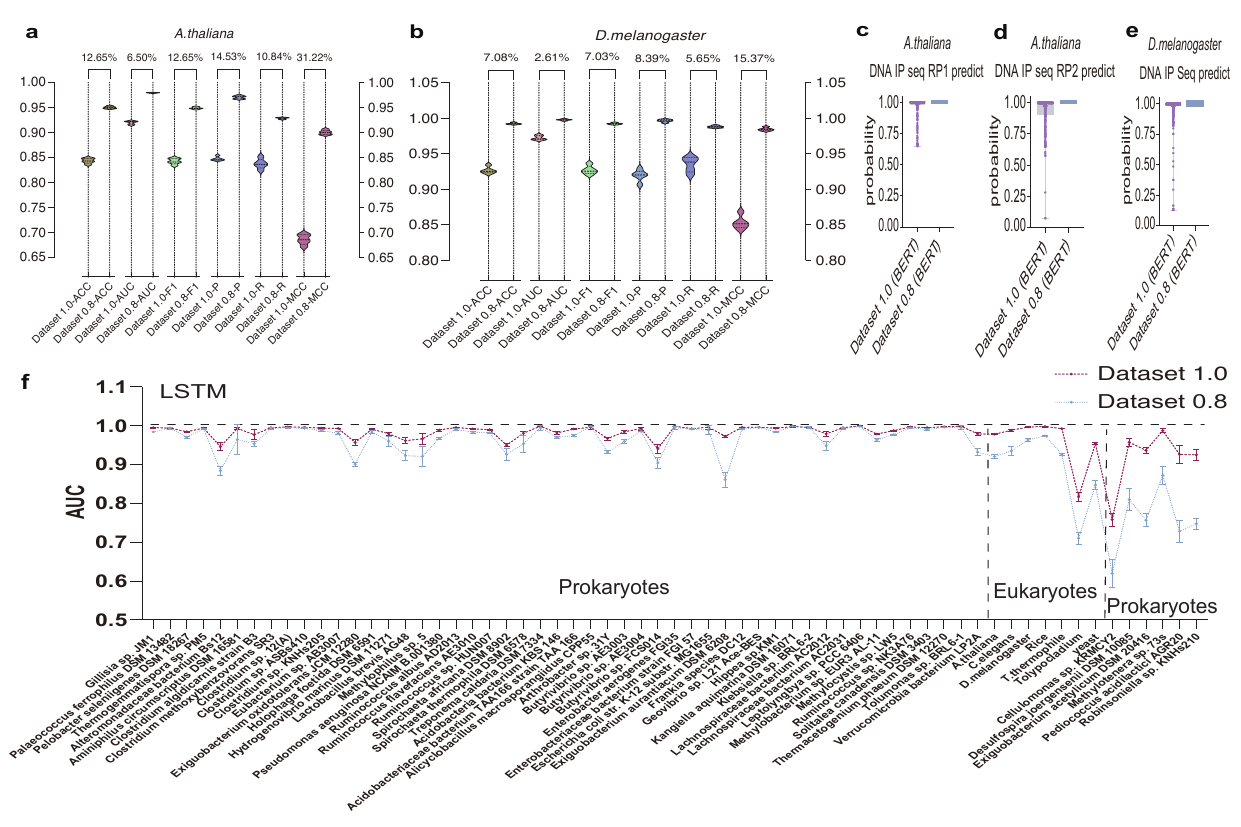}
	\caption{\textbf{ Performance enhancement and application validation through structural unit-based data optimization. }
	 (a–b) BERT performance based on the baseline dataset (Dataset 1.0) versus the optimized dataset (Dataset 0.8, constructed by filtering potential false negatives using structural units with $P(\text{pos}) \geq 0.80$  and rebalancing negative samples to achieve class equilibrium) is reported in A. thaliana and D. melanogaster. The metrics include ACC, AUC, F1, P, R, and MCC; relative improvements (\%) are labelled above the bars. (c–e) The probabilities of potential 6mA sites in independent 6mA-DNA-IP-Seq datasets are predicted: \textit{A. thaliana} replicates 1 (297 peaks) and 2 (376 peaks) and \textit{D. melanogaster} (801 peaks). (f) Cross-model and cross-species generalizability of ISAF4DNA under a raw 4-feature A/T/C/G input encoding, comparing the performance of the LSTM model based on Dataset 1.0 versus Dataset 0.8, evaluated by the AUC (see Supplementary Fig. 5 for additional metrics).
	} 
	\label{fig4} 
\end{figure}

Structural unit-guided data filtering significantly enhanced the discriminative performance of the model. With respect to the newly constructedDataset 0.8 ($P(\text{pos}) \geq 0.80$), BERT achieved relative improvements in ACC of 12.65\% in \textit{A. thaliana} and 7.08\% in \textit{D. melanogaster}, together with MCC gains of 31.22\% and 15.37\%, respectively (Fig. \ref{fig4}a–b). Consistent enhancements across other evaluation metrics, as shown in Fig. \ref{fig4}a–b, were reproducibly validated in Xoc BLS256 (Supplementary Fig. 4), which underscores the robustness of the augmentation strategy. To further evaluate its practical utility, we applied the model to independent 6mA-DNA-IP-Seq datasets with imprecisely localized methylation sites and included two biological replicates of \textit{A. thaliana} (RP1 and RP2) and one of \textit{D. melanogaster}. As shown in Fig. \ref{fig4}c–e, the models trained based on Dataset 0.8 consistently outperformed the baseline models across all three datasets (see Supplementary Tables 6–25 for detailed results). Collectively, these findings demonstrate that structural units not only enhance model performance but also reflect superior predictive ability based on independent experimental datasets that contain challenging methylation sites.

\subsubsection*{Generalizability across species and model architectures}

To validate the stable generalizability of ISAF4DNA and its model- and representation-agnostic nature, we evaluated 63 species across Bacteria, Archaea, and Eukarya using an LSTM architecture that is markedly different from BERT, with inputs reverted to the raw A/T/C/G alphabet. The predictive performance was systematically compared under structural unit pruned (P(pos)$\geq 0.80$) and unpruned conditions. As shown in Fig. \ref{fig4} and Supplementary Fig. 5, for most species, structural unit pruning significantly enhanced the key metrics (ACC, F1, AUC, and MCC); in limited cases (e.g., \textit{Ruminococcus} sp. NK3A76 and \textit{Solitalea canadensis} DSM 3403), the performance gain from cleaning was marginal because of near-ceiling performance. Overall, the consistent gains across models, species, and representations support a data-centric explanation—noise suppression via false-negative cleaning—rather than the results being artifacts of particular model architectures or input representations. This outcome establishes ISAF4DNA as a portable, testable decoding framework that is particularly effective in low-separability settings.

\subsubsection*{Database augmentation and experimental concordance}

By decoding 6mA across 63 species, our decoding framework(ISAF4DNA) not only consistently recovered known motifs catalogued in REBASE\cite{20}but also identified novel motifs (Supplementary Table SA1)(Folder A). For example, the GAGG motif, previously thought to be confined to the eukaryotic last common ancestor\cite{21}, is widespread in prokaryotes,  which indicates that its regulatory function predates eukaryotic divergence. Notably, conventional motif discovery tools such as MEME (Supplementary Fig. 6) also failed to reveal these patterns, highlighting the sensitivity of ISAF4DNA in capturing weak signals and evolutionarily conserved motifs.

In addition to reproducing and discovering motifs, ISAF4DNA also refines database annotations by resolving subtle recognition-site shifts. In Eubacterium sp. AB3007, the framework indicated positional shifts in the motifs CAAAA\underline{A} (CAAA\underline{A}A)and CAGA\underline{A}G (CAG\underline{A}AG) (Supplementary Files SA2-SA3). Although both motifs are catalogued in REBASE, their positional shifts had not previously been recognized (Supplementary Table SA1). Existing tools such as MEME are unable to resolve such subtle differences (Supplementary Fig. 7). Furthermore, the positional shift of CAAA\underline{A}A has not yet been -site shifts and suggest key candidates for fuobserved in vitro\cite{22}. These results provide critical bioinformatic evidence for targetture experimental testing.

Crucially, the ISAF4DNA findings not only support existing experimental findings but also provide higher-resolution insights. In T. thermophila, for instance, ISAF4DNA refined prior observations by confirming ApT(AT) as the “default core,” systematically quantifying both favourable and unfavourable effects of flanking bases and establishing a detailed hierarchy that ranks GC flanks above AT flanks (Supplementary Fig. 8 and Supplementary Table SA4). This outcome is fully consistent with prior experimental trends\cite{23} but provides a more granular interpretation. In E. coli, the GAT and ATC motifs identified by ISAF4DNA (Supplementary File SA5) exactly recapitulate the binding mode revealed by Dam methyltransferase crystal structures\cite{24}, offering a direct validation of computational prediction through structural biology.

Overall, ISAF4DNA bridges computational discovery, database refinement, and experimental validation, yielding a framework that not only decodes conserved regulatory grammar but also exposes subtle positional shifts and resolves fine-scale binding preferences. Such integrative power establishes a solid foundation for translational applications, from improving annotations in regulatory databases to informing the rational engineering of synthetic biology components and selecting key candidates for targeted experimental validation.

\subsection*{Evidence of language-like organizational patterns}
In cross-species analyses, the decoding framework identified two reproducible signatures of a language-like system—a derivational pathway of motifs (lexicon layer) and systematic deployment of redundancy (syntax layer).

\subsubsection*{Motif derivation pathway (lexicon layer)}

We observed and quantified a conserved derivation pathway across species, namely, AT → GAT/ATC → GATC, and the evidence is the following: (i) In T. thermophila, the early enrichment of GAT/ATC over AT, as quantified by P(pos) and AT-matched odds ratios (Supplementary Fig. 8a–b), indicated forward directionality. (ii) In prokaryotes, GATC was broadly conserved and , in some species, further extended to variants such as GGATCC, GATCC, or TGATC, while recognition of ancestral motifs (GAT, ATC, and GATC) persisted. This finding is consistent with functional expansion rather than replacement (Supplementary Fig. 9; Supplementary File SA6).
(iii) After standardizing species-specific sequence backgrounds using odds ratios (ORs) with a stringent threshold for strong effects (OR $\ge$ 2; 95\% confidence interval (95\% CI) of the log-OR excluding 0 after multiple testing correction), the critical steps in the derivation pathway "AT → GAT/ATC → GATC" displayed significant and consistent enrichment effects across the majority of species (15/19) (Supplementary Table SA7). These results provide strong, convergent statistical support for a broadly conserved derivation pathway across the species examined. 
(iv)Along the derivation pathway AT$\rightarrow$GAT/ATC$\rightarrow$GATC, the enrichment-based positive tendency\(P(\mathrm{pos}\mid m)\)  increased monotonically. For example, in T. thermophila, the submotifs GAT and ATC—derived from the ancestral motif AT (\(P(\mathrm{pos}\mid AT)\)=0.74)—showed higher \(P(\mathrm{pos}\mid m)\)  values (0.83 and 0.85, respectively)(Supplementary Table SA8). This trend was most pronounced for the derived GATC motif, where \(P(\mathrm{pos}\mid GATC)\) approached 1.00 in species such as E. coli . Biochemical measurements further indicated that E. coli EcoDam methylated the canonical GATC site at a rate of approximately three orders of magnitude faster than that for tested GATN variants (N $\ne$ C; e.g., GATG, GATT) \cite{25}. These findings support structural elaboration coupled with functional/kinetic enhancement and are consistent with previous predictions of the "GAT/ATC$\rightarrow$GATC" derivation pathway\cite{26}.
(v) Additional generative routes included recognition-site shifts (T\underline{T}\underline{A}A$\rightarrow$ \underline{A}AT\underline{T}), flank accretion (\underline{T}TA\underline{A}$\rightarrow$ A\underline{T}TA\underline{A}T; a plausible physical route) (Supplementary Fig. 10), and repetitive assemblies (e.g., GATCGATC) in some taxa (Supplementary Table A9). 
(vi) As a parallel line of evidence in multicellular eukaryotes, we identified a GA-centred track that increases in complexity via flank accretion/replacement (Supplementary Fig. 3b-c). This pattern showed strong cross-validation between \textit{A. thaliana} and \textit{D. melanogaster} (Supplementary Fig. 3a), indicating the portability of the derivation logic across multicellular eukaryotes. Together, structural continuity, functional optimization, and multiple generative mechanisms support a derivation pathway based on evolutionary patterns that is applicable across both prokaryotes and eukaryotes.

\subsubsection*{Redundancy types and deployment preferences (syntax layer)}

Using structured data from 63 species (Supplementary Folders B, C), we identified and characterized two reproducible redundancy patterns in 6mA information organization: (i) an anchor-type pattern, where selective flanking biases around a conserved core motif allow flanking regulatory flexibility (Supplementary Fig. 2c); and (ii) fuzzy-type clusters composed of distributed rule units with greater positional tolerance, indicating stronger redundancy-based robustness (Supplementary Fig. 2a–b). Anchor-type patterns are more frequent in prokaryotes (33/56, 58.9\%; Supplementary File SA11), whereas fuzzy-type clusters predominate in multicellular systems (4/4)(Supplementary Folder C). The two modes can co-occur—for example, yeast shows both an AGGT(A/T) anchor-type signal and an adenine-enriched fuzzy-type cluster (Supplementary File SA10). We systematically catalogued the flanking preferences and distributions of the anchor-type structures (Supplementary Files SA11–SA12). The inclusion of the phylogenetic context revealed the continuity of these flanking biases along related branches (Supplementary Fig. 10), which is consistent with functional relevance and cross-lineage conservation. Overall, the deployment and flanking patterns indicate that DNA regulation extends beyond a single “core motif” to context-dependent, nested architectures that confer robust recognition and adaptive function.

\subsection*{EpigenoLinguistics: A multilayer language-like framework for DNA regulation}
Using our unsupervised decoding approach, we identified redundant nested architectures at the syntactic layer and stable motifs with their derivational pathways at the lexical layer that are conserved across species. Their deployment patterns and context dependencies exhibited reproducible regularities, which indicate an organizational logic that extends into the pragmatic layer. On this basis, we propose EpigenoLinguistics, a preliminary, empirically grounded framework for a language-like system of DNA regulation. Drawing on the multilevel analytic tradition of linguistics, this framework provides a structured account of DNA’s coding logic—from the composition and evolution of regulatory motifs (lexicon) through their combinatorial rules (syntax) to system-level deployment strategies (pragmatics).

\begin{figure}[H]
	\centering
	\includegraphics[width=1.0\textwidth]{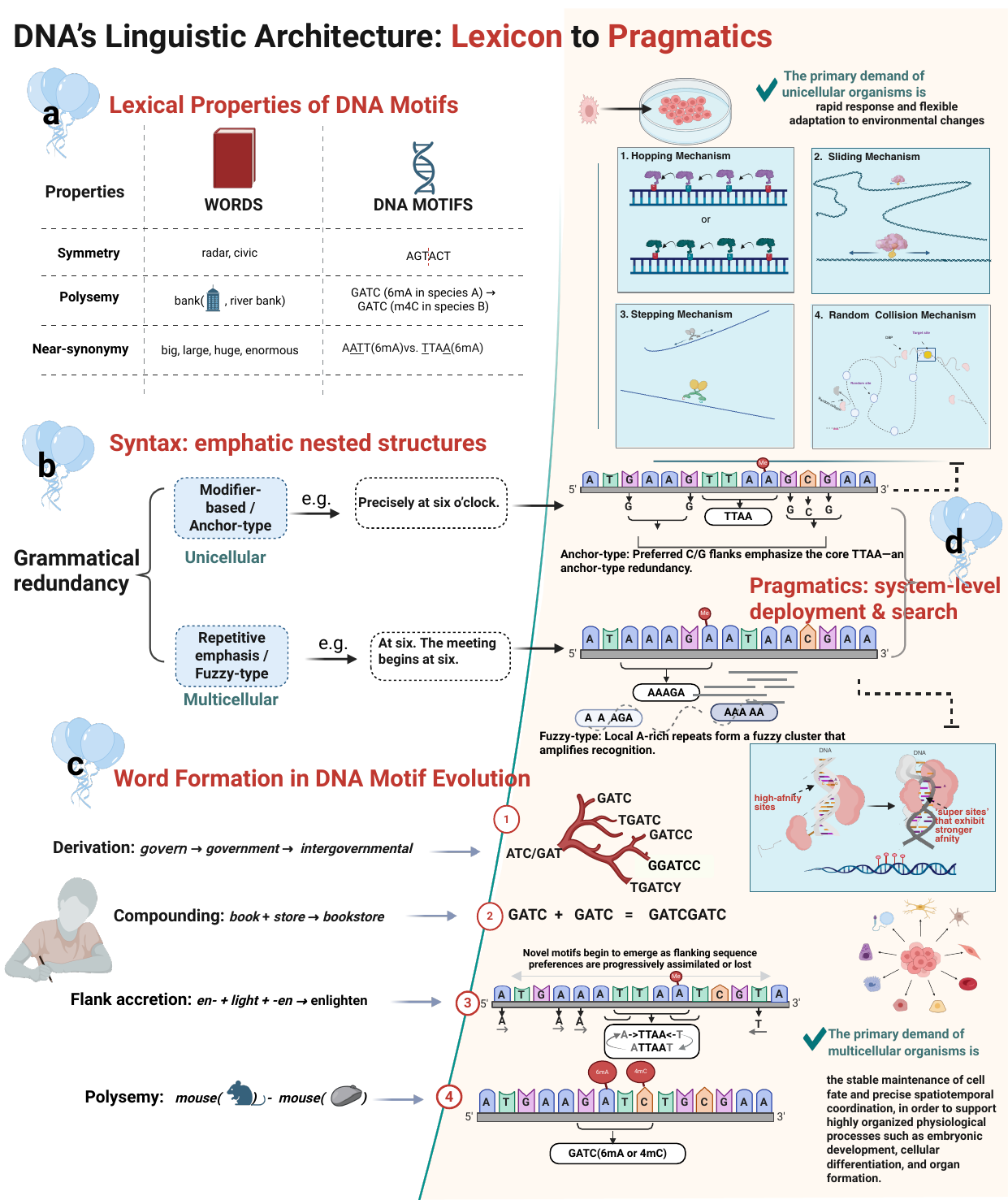}
	\caption{\textbf{EpigenoLinguistics: DNA’s linguistic architecture from lexicon to pragmatics.}
		 (a) Lexicon—stable, recognizable motifs that act as regulatory “word” items and display lexical properties (e.g., symmetric structure, polysemy, and near-synonymy). (b) Syntax—emphatic nested redundancy: modifier-based/anchor-type vs. repetitive-emphasis/fuzzy-type (the examples illustrate C/G flanks that emphasize TTAA and local A-rich fuzzy-type clusters). (c) Word formation in motif evolution—derivation (GAT/ATC→GATC, with extension to GGATCC), compounding (GATC+GATC→GATCGATC), flank accretion (core-preserving extension, e.g., TTAA→ATTAAT), and polysemy across modifications/context. (d) Pragmatics—system-level deployment and search (unicellular systems tend to deploy anchor-type redundancy; multicellular systems favour fuzzy clusters).
	} 
	\label{fig5} 
\end{figure}

By analogizing DNA regulatory motifs to words in natural language, we revealed word-like structural properties (Fig. \ref{fig5}a) and evolutionary processes that resemble word formation (Fig. \ref{fig5}c). At the lexicon layer, stable, recognizable motifs function as regulatory lexical items (e.g., GATC, AATT). Although their forms and core roles are conserved, their functional meanings can differ—for example, the GATC motif is associated with 6mA modification in one species but with 4mC(N4-methylcytosine) in another(Supplementary Table SA13). As described in the preceding section, new functional units arise through processes of structural derivation or motif composition, which parallel the morphological derivation and compounding found in natural language (Fig. \ref{fig5}c). Thus, these DNA “words” constitute composable, interpretable, and derivable information-bearing units \cite{27}. In linguistics, a word is defined as the minimal unit that integrates form, meaning, and function. Words can exhibit polysemy and semantic shifts, and they anchor syntactic and semantic organization \cite{28,29}. Accordingly, we define stable, identifiable, and combinable/derivable motifs as regulatory lexical items whose semantics are instantiated by modification type or context-dependent functional labels.

Redundancy is not excess but a cognitive–pragmatic mechanism that improves clarity and fault tolerance. In natural language, it typically appears as repetition and intensification that maintains coherence, marks focus, and supports comprehension in complex contexts \cite{30}. At the grammar–semantics interface of DNA regulation, we categorize two primary forms of redundancy (Fig. \ref{fig5}b). (i) Anchor-type (modifier-based) redundancy features a conserved core motif flanked by preferential modifiers (e.g., TTAA(C/G)). This nested "core + modifiers" scheme is functionally analogous to the intensifying effect of adverbs such as “precisely” in natural language and serves to amplify the core signal. (ii) Fuzzy-type redundancy clusters are formed by the localized stacking of short, functionally convergent sequence units with diffuse boundaries (e.g., A-rich tracts). This architecture is comparable to repetitive emphatic constructions (e.g., sentences such as "At six. The meeting begins at six."). The "parallel" arrangement of multiple units enhances recognition specificity and system-level fault tolerance. Although mechanistically distinct, both share an isomorphic design—nesting plus repetition—that amplifies regulatory signals and tunes them to context, forming emphatic nested structures within the DNA grammar. These patterns increase the expressive efficiency, positional precision, and system robustness without altering the core semantics.

Across biological systems, grammatical redundancy is deployed in ways that are compatible with system-level pragmatic demands (Fig. \ref{fig5}d). Unicellular organisms preferentially employ anchor-type redundancy (a conserved core with flanking preferences), which helps prioritize site recognition under frequent and unpredictable environmental stimuli. The flanking context can modulate local DNA conformation\cite{31}and the susceptibility of target-base flipping\cite{32,33,34}, thereby tuning enzyme binding and catalytic timing\cite{35}; this is consistent with established target-search modes—1D sliding, hopping/intersegment transfer, and 3D diffusion\cite{36,37,38,39,40,41}.  Multicellular organisms more often exhibit fuzzy-type redundancy clusters—spatially coordinated modules composed of partially nonidentical yet functionally convergent elements—which can be explained by increased short-timescale facilitated rebinding (effective local rebinding probability)\cite{42,43}, DNA-shaped readouts\cite{44}, and multiprotein cooperativity\cite{45}that jointly form local high-affinity “supersites”\cite{46}that suit larger and more structured genomes.

\section*{Discussion}


We present an interpretable and mechanism-transparent AI framework, ISAF4DNA, which employs state-aware encoding and structural-semantic decoding to overcome the limitations of black-box prediction and context neglect, thereby establishing an operable pathway for deciphering DNA regulatory information. Using DNA 6mA prediction as a case study, our framework not only achieves steady improvements in model performance but also reveals a language-like architecture of DNA information: lexical units (motifs), redundant nested syntax, and context-dependent semantic reinforcement. Building on these findings, we advance EpigenoLinguistics, which considers DNA a dynamic regulatory language rather than a static sequence, thereby closing the theoretical gap in the application of language models to biological sequences. We further complete and correct existing resources by identifying new motifs and positional shifts to achieve the superior detection of weak signals and fine-grained site localization. We also provide reusable, program-ready design templates to establish a mutually reinforcing AI$\leftrightarrow$experiment cycle.

Viewed through a systems lens, 1D sequence redundancy and 3D chromatin folding likely cooperate to form a layered epigenetic control system: redundancy offers a composable, error-tolerant, reversible code, whereas folding endows modification states with conformational stability \cite{47}. External cues can engage TLR9-linked inflammation and elicit DNA damage repair (DDR), resulting in DNA damage and subsequent repair \cite{48}—events that may locally reset the sequence context and epigenetic marks and alter information writing and retention. In \textit{A. thaliana}, 6mA and 5mC are concentrated in centromere-proximal heterochromatin \cite{21,49}, which is consistent with the epigenetic stabilization of 3D architecture. Together with the write/erase nature of these marks, this result points to a read–write channel between 1D redundancy and 3D chromatin, enabling a balance between rapid response and steady-state maintenance.

Redundancy is a structured—not random—organizing device that in information theory and neuroscience, underwrites noise robustness and steady-state maintenance\cite{50,51}. Analogously, in DNA regulation, the observed anchor-type and fuzzy-type redundancies both stabilize modification outputs under changing conditions and, via structural rewiring and contextual allocation, afford tuneable strength and error tolerance. Our evidence shows multilayer organization—recursive nesting (syntactic redundancy), combinatorial formation (lexical derivation), and context dependence (flank/neighbourhood reinforcement)—that formally echoes the generalizability and context sensitivity of human language. Such isomorphism provides an independent lens for Chomsky’s LAD as an innately specified generative system \cite{52} and implies that human language may reflect a redeployed biological coding logic rather than being a purely de novo cultural artifact.

To our knowledge, this is the first systematic empirical support for the multilayered, language-like regulatory organization of DNA obtained at base-pair (sequence-level) resolution with an interpretable AI structural–semantic decoding pipeline. Spanning 63 species, these multilayer patterns motivate EpigenoLinguistics, which reframes “DNA as language” from metaphor to mechanism-linked function. More generally, our explanation-first strategy embeds interpretability and testability in the model objective and redirects AI from accuracy gains to mechanism discovery and theory building. The resulting template is applicable to AI4Science (symbolic–neural integration, experiment-aware active learning, and mechanism metrics) and, in principle, generalizes beyond 6mA to 4mC, 5mC, and RNA modifications. Prospective tests—cross-modification transfer, external dataset challenges, and targeted perturbations—will consolidate this read–write–design pathway for regulatory coding systems.

\section*{Methods}

\paragraph{Datasets and preprocessing}
We focused on DNA N6-methyladenine (6mA) as the target for decoding. 6mA has been reported across diverse species\cite{53,54} and exhibits pronounced context dependence and regulatory complexity\cite{55,56}, which makes it an appropriate model system for investigating DNA regulatory grammar and its language-like properties. To systematically evaluate the feasibility and cross-species and cross-model applicability of the ISAF4DNA decoding framework, we constructed the dataset in two stages. First, we validated the method in three representative species, and second, we expanded it to 63 species to enable generalized analyses of structural grammar and evolutionary inference. All of the data were derived from single-molecule real-time (SMRT) sequencing.

We selected \textit{A. thaliana}, \textit{D. melanogaster}, and \textit{Xoc BLS256} to represent plants, animals, and microorganisms, respectively. Positive samples were obtained from MethSMRT \cite{57} and the published literature \cite{58}, with a threshold of \( \mathrm{QV} \geq 30 \), and were deduplicated using CD-HIT at an 80\% sequence-similarity threshold. The negative samples were randomly drawn from the reference genome at unmethylated A sites; 41-bp sequence windows were constructed (with the target methylated A located at position 21) and were required to be >20 bp from any known 6mA site. For each species, 20\% of the positive samples were reserved as a hold-out set to support cross-species cross-validation evaluation (Supplementary Fig. 3a).

\noindent\textbf{Extended dataset (63 species)}. We expanded the coverage to 56 prokaryotic and 7 eukaryotic species. Positive sites were likewise constructed from MethSMRT or peer-reviewed literature \cite{59,60}, and the preprocessing pipeline was identical to that used for the representative species above. To prioritize data volume, QV thresholds were adjusted for some species (specific thresholds and sample sizes are provided in Supplementary Table SA14).

\noindent\textbf{Phylogenetic analysis.}  A phylogenetic analysis was conducted on 16S/18S rRNA by consulting public databases and performed using MEGA\cite{61}.

\paragraph{State-Aware Symbolic Encoding System}

We view DNA as a time-varying information system: during enzyme–DNA recognition, observable transitions occur between the unbound duplex and binding-related states \cite{24,62,63,64}. To operationalize this view, we define a 41-bp window centred on each candidate site (centre index p=21). Let the nucleotide set be B={A,T,C,G} and the relative position index be P={1, …, 41}. Their Cartesian product S=B×P yields 164 discrete joint states, where each element (b, p) denotes “base b at relative position p.” We map every joint state to a fixed token to form a state-aware, high-entropy symbolic representation in which the relative position acts as an observable proxy for the local conformation/microenvironment.
Compared with that of traditional A, T, C, and G encoding (maximum of 2 bits/symbol), this encoding has a theoretical maximum entropy of \(H_{\max}(X)=\log_{2} n\)\cite{50}.
Letting \(n=|S|=164\) gives \(H_{\max}=\log_{2}|S|=\log_{2}164
\approx 7.36\,\text{bits/symbol}\), thereby increasing per-site information capacity and contextual resolution (Fig. \ref{fig1}a).
To assess robustness, we kept data splits and hyperparameters fixed and varied only the state-to-token mapping (dictionaries dict1–dict3) for controlled comparisons. For visualization, the 164 tokens were embedded into a common vector space and projected with principal component analysis (PCA; Fig. \ref{fig2}d). This encoding provides a stable and testable input basis for the downstream identification of syntactic (grammar-like) structures.

\paragraph{Neural network training and evaluation}

To evaluate the applicability and predictive performance of the state-aware symbolic encoding for DNA semantic recognition, we used BERT as the backbone model \cite{65} and designed comparative experiments with two input representations: the raw DNA alphabet (A/T/C/G; 4 features) and the state-aware symbolic sequence (164 features) (Supplementary Fig. 1). All of the experiments were conducted based on Dataset 1.0, with the data partitioned into training (80\%) and test (20\%) sets. We used the following metrics to evaluate the model performance: the receiver operating characteristic (ROC) and the area under the ROC curve (AUC), accuracy (ACC), precision (P), recall (R), F1, and the Matthews correlation coefficient (MCC). These metrics are defined as follows:
\begin{equation}
	P = \frac{TP}{TP + FP}
\end{equation}

\begin{equation}
	R = \frac{TP}{TP + FN}
\end{equation}

\begin{equation}
	\frac{1}{F_1} = \frac{1}{2} \left( \frac{1}{P} + \frac{1}{R} \right)
\end{equation}

\begin{equation}
	ACC = \frac{TP + TN}{TP + TN + FN + FP}
\end{equation}

\begin{equation}
	MCC = \frac{TP \cdot TN - FP \cdot FN}{\sqrt{(TP + FN)(TP + FP)(TN + FP)(TN + FN)}}
\end{equation}

where TP, TN, FP, and FN represent true positive, true negative, false positive, and false negative, respectively. The corresponding performance plots are shown in Fig. \ref{fig2}a–b. Furthermore, to support the interpretability of ISAF4DNA, we performed cross-validation, which provided auxiliary evidence for structural-grammar decoding (Supplementary Fig. 3a).

\paragraph{Structural unit mining and sequence annotation}

To identify the putative regulatory structures, we applied the Apriori algorithm\cite{66} to the state-aware symbolic sequences from the positive samples and extracted combinations occurring in  $\ge 6\%$ of sequences (minimum support 0.06) as candidate structural units. Unit specificity across classes was quantified as
\begin{equation}
P(\mathrm{pos}\mid m)
= \frac{f_{\mathrm{pos}}(m)}{f_{\mathrm{pos}}(m)+f_{\mathrm{neg}}(m)}\,
\end{equation}

where \( f_{\text{pos}}(m) \) and \( f_{\text{neg}}(m) \)  are the number of sequences in the positive and negative sets, respectively, that contain the motif \( m \) (at most one count per sequence). High-confidence units were retained for downstream cleaning and validation (see Supplementary Tables 3–5).

After a WebLogo analysis\cite{67} and the structure-level visualization/annotation of positive sequences from \textit{A. thaliana}, \textit{D. melanogaster}, and Xoc BLS256 (Supplementary Fig. 2), we identified the representative redundant structures. We defined redundancy as the repetition, overlap, or nesting of grammatical units within a local window. Based on the spatial distribution, we grouped these clusters into two patterns: (i) anchor-type clusters that feature a stable core with selective flank preferences (Supplementary Fig. 2c) and (ii) fuzzy-type clusters that comprise multiple distributed rule units with indeterminate boundaries and a diffuse layout (Supplementary Fig. 2a–b). To evaluate the core–flank combinatorial preferences, we devised a layered support metric (Methods-Data analysis; results in Supplementary Files SA11–SA12), which quantified the occurrence and enrichment of the individual flanks and their combinations.

\noindent\textbf{completion procedure.}  Because structural unit mining may miss low-support yet highly specific structures, we implemented a completion step to improve the coverage of ISAF4DNA-based decoding. We first collected positive sequences that did not contain any high-confidence units ($P(\text{pos}) \geq 0.99$) to form a set of unannotated segments. We then rescanned this set with the Apriori algorithm under relaxed support but stricter specificity thresholds and recovered the low-frequency, highly discriminative combinations, which were incorporated into the structural unit library (Supplementary Folder B). This completion step increases the depth and completeness of the rule library at the species-level sequence resolution (Supplementary Table SA1).

Although high-frequency structures are enriched in the positive samples, their actual contribution to semantic discrimination remains to be established. Accordingly, we designed a semantic gradient-based cleaning experiment. Using the specificity score $P(\mathrm{pos}\mid m)$, we constructed five sets of high-confidence structural units at thresholds $\tau \in \{0.95,\,0.90,\,0.85,\,0.80,\,0.75\}$. For each $\tau$, we scanned the original negative set: any sequence that contained at least one structural unit m with $P(\mathrm{pos}\mid m)$ $\ge$ $\tau$ was labelled a potential false negative and removed.

For each threshold, we also created a random-deletion control by uniformly removing the same number of negative sequences as in the cleaned set to keep the class sizes matched. The models were then retrained and evaluated based on the cleaned datasets, and the remaining counts of positive and negative samples were recorded; the results are shown in Fig. \ref{fig3}c.
\paragraph{Data Augmentation and Application}

To quantify the practical benefits of the structural units for data quality and model generalizability, we constructed a structure-guided cleaned dataset, Dataset 0.8. Specifically, during negative sampling only, we removed any sequence that contained at least one structural unit with $P(\mathrm{pos}\mid m)$ $\ge$ 0.80; the positive samples were left unchanged. Using a fixed BERT architecture and identical hyperparameters, we trained the models based on Dataset 1.0 and Dataset 0.8 and compared their performance on the 6mA sequence classification task (Fig. 4a–b; Supplementary Fig. 4). Performance is reported using ACC, AUC, F1, P, R, and MCC.

To evaluate the application performance, we introduced two independent 6mA-DNA-IP-Seq datasets \cite{68}: A. thaliana (two replicates) and D. melanogaster. Given that 6mA-DNA-IP-Seq does not provide single-nucleotide resolution, all A positions within ±50 bp of each peak centre were treated as prediction targets. BERT models trained separately based on Dataset 1.0 and Dataset 0.8 were used to score the A sites on the forward and reverse strands. Within each enriched peak, the highest-scoring A site was regarded as the candidate methylation site. A visualization is provided in Fig. 4c–e; see Supplementary Tables 6–25 for further details.

\paragraph{Cross-Species, Cross-Model Generalizability and Structural Robustness}

We replicated the analysis in an LSTM architecture \cite{69} (distinct from BERT’s transformer architecture) and reverted the inputs to a four-feature (A/T/C/G) one-hot encoding to test the model- and representation-agnostic nature of ISAF4DNA. Using the 63-species dataset 1.0 as the baseline, we applied the same rule to remove—from the negative pool only—any sequence containing a structural unit with $P(\mathrm{pos}\mid m)$ $\ge$ 0.80, yielding the cleaned dataset 0.8 (positives unchanged). For each species, the LSTM models were trained separately based on Dataset 1.0 and Dataset 0.8 and evaluated using a five fold cross-validation with matched splits (Fig. \ref{fig4}f; Supplementary Fig. 5).

\paragraph{Statistical analysis}

\textbf{Flank preferences and effect size characterization.}From the high-confidence structural units identified by Apriori frequent-itemset mining, we observed that some core motifs exhibited systematic preferences for particular flanks (Supplementary Folder B). To quantify this “core–flank” preference, for each core motif m, we compared the distributions of A/T-group versus C/G-group flanks among the positive samples using support as the primary measure. Statistical inference was performed with Welch’s t tests; we report the standardized effect size (\( \eta^2 \)), mean differences with 95\% confidence intervals (95\% CI), and P values (Supplementary File SA11).  This “preference” captures the selective pairing tendencies of the flanks, which may modulate recognition and catalytic timing by altering local flexibility and binding affinity\cite{35}. To avoid ad- hoc cutoffs, we graded the effects by quartiles of the \( \eta^2 \) distribution computed from 56 prokaryotic species with comparable flanks: strong (\(\eta^2 \ge P_{75}\)), medium (\(P_{50} \le \eta^2 < P_{75}\)), weak (\(P_{25} \le \eta^2 < P_{50}\)), and none (\(\eta^2 < P_{25}\) ). Box-and-whisker plots summarize the per-species and overall distributions for visual assessment (Supplementary File SA12). In downstream analyses, we prioritized the flanks graded as medium or above (\(\eta^2\) $\ge$ P50) that also passed multimetric screening(Supplementary Files SA11, SA12).

\textbf{Conditional enrichment for derivational edges} To assess whether adding a child motif confers additional discriminative power conditional on the presence of its parent motif, we computed the odds ratio (OR) as in \cite{70}:
\begin{equation}
\mathrm{OR}(m)=\frac{a/c}{b/d}=\frac{ad}{bc}
\end{equation}
where \(a=f_{\mathrm{pos}}(m)\) and \(b=f_{\mathrm{neg}}(m)\) are the counts of motif \(m\) in the positive and negative sets, respectively; \(N_{\mathrm{pos}}\) and \(N_{\mathrm{neg}}\) are the counts of the parent motif in the positive and negative sets; and \(c=N_{\mathrm{pos}}-a\), \(d=N_{\mathrm{neg}}-b\).
All inference was performed on the log scale:
\begin{equation}
\log\mathrm{OR}=\ln\!\big(\mathrm{OR}(m)\big), 
\mathrm{SE}=\sqrt{\frac{1}{a}+\frac{1}{b}+\frac{1}{c}+\frac{1}{d}}.
\end{equation}
The Wald 95\% confidence interval on the log scale is
\begin{equation}
\log\mathrm{OR}\ \pm\ 1.96\,\mathrm{SE},
\end{equation}
and can be exponentiated to obtain the interval for the \(\mathrm{OR}\).

When any of \(a,b,c,\) or \(d\) is small (e.g., \(\le 5\)) or zero, we apply the Haldane--Anscombe continuity correction:a'=a+0.5, b'=b+0.5, c'=c+0.5, d'=d+0.5,
and recompute\(\mathrm{OR}^*=\frac{a'd'}{b'c'}\),
\(\log\mathrm{OR}^*=\ln\!\big(\mathrm{OR}^*\big)\),
\(\mathrm{SE}^*=\sqrt{\frac{1}{a'}+\frac{1}{b'}+\frac{1}{c'}+\frac{1}{d'}}\).The corresponding 95\% confidence interval is \(\log\mathrm{OR}^*\ \pm\ 1.96\,\mathrm{SE}^*\).

\textbf{Criterion and reporting.}We classified a single parent-to-child derivative edge as having a strong effect if the OR was$\ge$2 and its 95\% confidence interval on the log scale did not include 0. The results are reported in Supplementary File SA8.

\bibliography{references}
\section*{Acknowledgments}
This work was supported by the Major Project of the National Social Science Fund of China (Project No. 21\&ZD331), which enabled the development of this research. We would like to express our sincere gratitude to the funding agency for their trust and support. We also thank all colleagues for their invaluable expertise and assistance throughout the course of this project.
\section*{Author contributions statement}

Li Yang and Dongbo Wang conceived and supervised the project. Li Yang designed the methods, conducted the experiments, and wrote the manuscript. Dongbo Wang reviewed and revised the manuscript. All authors approved the manuscript.

\section*{Additional information}

\textbf{Data availability}

All key datasets supporting the findings of this study have been submitted directly to the Nature submission system as supplementary data files. These datasets are available to editors and reviewers during peer review. Upon publication, the complete datasets will be made publicly available via a dedicated GitHub repository.

\textbf{Competing interests}
The authors declare no competing interests.

\end{document}